\documentclass[pra,twocolumn,a4paper]{revtex4-1}
\usepackage{amssymb}
\usepackage{epsfig}
\usepackage{amsmath}
\usepackage{graphicx}
\usepackage{bm}
\usepackage{times}
\usepackage{color}

\newcommand{\bn}[1]{\langle{#1}\rangle}
\newcommand{\Dc}{\Delta}

\begin{document}

\title{ Optomechanics with molecules in a strongly pumped ring cavity}
\author{R. J. Schulze, C. Genes and H. Ritsch}
\affiliation{Institute for Theoretical Physics, University of
Innsbruck, and Institute for Quantum Optics and Quantum Information,
Austrian Academy of Sciences, Technikerstrasse 25, A-6020 Innsbruck,
Austria}
\date{\today }

\begin{abstract}
Cavity cooling of an atom works best on a cyclic optical transition
in the strong coupling regime near resonance, where small cavity
photon numbers suffice for trapping and cooling. Due to the absence
of closed transitions a straightforward application to molecules
fails: optical pumping can lead the particle into uncoupled states.
An alternative operation in the far off-resonant regime generates
only very slow cooling due to the reduced field-molecule coupling.
We predict to overcome this by using a strongly driven ring-cavity
operated in the sideband cooling regime. As in the optomechanical
setups one takes advantage of a collectively enhanced field-molecule
coupling strength using a large photon number. A linearized
analytical treatment confirmed by full numerical quantum simulations
predicts fast cooling despite the off-resonant small single molecule
- single photon coupling. Even ground state cooling can be obtained
by tuning the cavity field close to the Anti-stokes sideband for
sufficiently high trapping frequency. Numerical simulations show
quantum jumps of the molecules between the lowest two trapping
levels, which can be be directly and continuously monitored via
scattered light intensity detection.
\end{abstract}

\maketitle

\section{Introduction}

In recent years cavity cooling and manipulation of small polarizable
particles has developed into an extended field of
research~\cite{cavcoolrev,Lev-prospects-08}. Considerable
theoretical progress understanding the strongly coupled atom-field
dynamics including particle motion has been made in the past
decades~\cite{ritsch97, Morigi2007a, deachapunya-slow-07,
salzburger2009collective}, and has been accompanied by impressive
experimental achievements~\cite{murr2006three,zhang2009experimental,
leibrandt2009cavity,slama2007superradiant,
hemmerich06,brennecke2008cavity}.

As recently discussed in a detailed review
article~\cite{Lev-prospects-08}, a generalization to molecule
cooling is limited by the inability of achieving sufficiently strong
dispersive coupling to the cavity field. At the usable frequencies
far from any internal molecular resonance the ratio of
polarizability and mass is orders of magnitude below the value close
to resonance. Although cooling persists in principle it slows down
too much to be useful. One way out of this dilemma is the use of
collective enhancement using a high density of particles inside the
mode volume~\cite{salzburger2009collective}. Again in principle the
mechanism has been shown to work for pre-cooled
atoms~\cite{chan2003observation}, but it turns out to be technically
challenging to implement the required high pressure and low
temperature beam sources for molecules~\cite{deachapunya-slow-07}.

As an alternative route one can instead take a clue from the field
of optomechanics, where motional degrees of freedom of large
vibrating objects such as
micro-mirrors~\cite{Groeblacher-demonstration-09}, toroidal
microresonators wall vibrations~\cite{Schliesser-resolved-09},
nano-membranes~\cite{thompson-strong-08}, levitated
nanospheres~\cite{chang-cavity-09} or even
viruses~\cite{isart-towards-09} are manipulated via radiation
pressure coupling to a light field. Such systems not only exploit
directed collective scattering (reflection) from a large number of
very weak scatterers but in addition rely on generating a stronger
light force just by applying very high light intensities. As the
effective mechanical coupling grows with the intracavity field
amplitude (i.e. the square root of the photon number), the
inherently small single photon coupling rate, proportional to the
extent of the objects zero-point-motion, can be augmented by several
orders of magnitude to produce strong friction forces.

\begin{figure}[t]
\centerline{\includegraphics[width=0.36\textwidth]{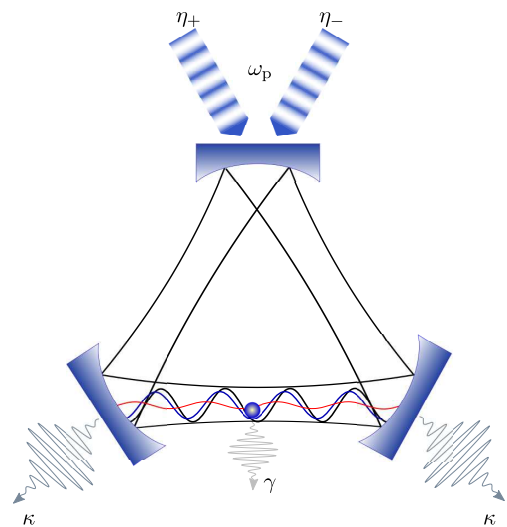}}
\caption{A ring cavity with two degenerate modes (cosine: black line
and sine: red line) with the same wave vector $k$. The cosine mode
is strongly pumped and is responsible for trapping a polarizable
particle around one of its intensity minima. Depicted in the figure
is also the effective potential seen by the particle (blue line) as
it is shifted around depending on the dynamical occupancy of the
sine mode.} \label{setup}
\end{figure}

In this work we consider the corresponding limit for a weakly
polarizable particle (molecule) in a ring cavity. In a previous
analysis concentrating in the low photon number but strong
atom-field coupling limit the ring cavity geometry has been shown to
lead to enhanced cooling of atoms by only a factor two at
best~\cite{gangl-cold-00,murr-large-06}. Let us instead focus on the
high-field limit for a symmetrically pumped ring cavity as in
typical optomechanical systems to study the corresponding coupling
enhancement for a very weakly coupled molecule (see
Fig.~\ref{setup}). For symmetric pumping it proves useful to
decompose the ring cavity field in a cosine and sine mode instead of
the more common propagating mode expansion. In this basis photons
from the highly excited driven cosine mode are scattered into the
almost empty sine mode. Amplitude and phase of this weak single
molecule light scattering to the sine mode is highly spatially
dependent and strongly enhanced by the presence of a large field
amplitude in the cosine mode. The situation for a high field seeking
particle is illustrated in Fig.~\ref{setup}. The particle is trapped
close to the antinodes of the highly populated cosine mode, while
the $\pi/2$-shifted sine mode provides the weakly--populated quantum
mode that produces sideband cooling. Actually an analogous situation
will occur for any cavity geometry with two almost degenerate modes
with differing mode functions.

In the following using different theoretical approaches we will
demonstrate two central physical effects for such a setup: i) the
efficiency of cooling in the dynamical regime is considerably
improved over cooling in a typical standing wave cavity with large
circulating power and ii) for particles trapped in a standing wave cosine mode in a ring cavity
the coupling of the single molecule to the corresponding sine mode is equivalent to an
optomechanical coupling with a coupling strength enhanced by a power
of the intracavity cosine mode photon number.

The paper is structured as follows: in Sec. II we present an
effective quantum Langevin description for the coupled dynamics of a
polarizable particle moving in the field of a ring cavity. In Sec.
III we ignore noise terms and study important dynamical features of
the underlying classical dynamics and compare this to the typical
standing wave cavity cooling. In Sec. IV we concentrate on the case
of an intense pump field with strongly confined particle motion,
where the dominant interaction couples the sine mode to the particle
motion. In Sec. V we numerically study the cooling process using
wave function simulations and we present numerical examples for
quantum trajectories exhibiting the evolution of the molecule's
state under continuous monitoring of the scattered field. Sec. VI
contains more detailed discussions on the physical interpretation of
the cooling mechanism in this setup. Conclusions are presented in
the closing Sec. VII.

\section{Quantum Langevin approach to ring cavity cooling}

We consider a one-dimensional configuration as illustrated in
Fig.~\ref{setup}, where a polarizable particle of mass $m$ modeled
as a two-level system with energy splitting $\hbar \omega _{a}$ is
off-resonantly coupled to the field in a ring cavity. For a wave
vector $k$ (corresponding frequency $\omega _{c}$) there are two
degenerate counter-propagating modes of the cavity driven with
amplitudes $\eta _{+}=\eta _{-}=\eta $. One can rethink this in
terms of cosine and sine standing wave modes, with the cosine mode
driven with $\eta $ while the sine mode is unpumped. The two optical
modes are represented by field operators $a_{c}$ (cosine mode), with
commutator $\left[ a_{c},a_{c}^{\dagger }\right] =1$ and $a$ (sine
mode) with $\left[ a,a^{\dagger }\right] =1$. The cosine mode is
driven by a laser at frequency $\omega _{p}$ and amplitude $\eta$
with detuning $\Delta =\omega _{p}-\omega _{c}$. Both fields decay
at rate $2\kappa $ through the mirrors. The particle's quantized
position and momentum operators are $\hat{x}$ and $\hat{p}$, $\left[ \hat{x},%
\hat{p}\right] =i\hbar $. The field, strongly detuned from the
particle's resonance by $\Delta _{a}=\omega _{a}-\omega _{p}$, is
coupled with strength $g$ so that we have very low saturation and
the spontaneous emission $\gamma$ plays only a minor role, as
discussed in more detail later in Sec.~\ref{discussion}. The
position dependent intracavity intensity seen by a particle at
position $\hat{x}$ can be
written as $(a_{c}^{\dagger }\cos k\hat{x}+a_{s}^{\dagger }\sin k\hat{x}%
)(a_{c}\cos k\hat{x}+a_{s}\sin k\hat{x})$. Elimination of the internal
degrees of freedom of the particle in the limit of very large detuning $%
\Delta _{a}\gg \gamma $ leads to an effective position modulated
energy shift~\cite{gangl-cold-00}
\begin{equation*}
U(\hat{x})=a_{c}^{\dagger }a_{c}U_{c}\left( \hat{x}\right) +a_{s}^{\dagger
}a_{s}U_{s}(\hat{x})+\left( a_{c}^{\dagger }a_{s}+a_{c}a_{s}^{\dagger
}\right) U_{cs}(\hat{x}),
\end{equation*}%
where $U_{c}\left( \hat{x}\right) =U_{0}\cos ^{2}k\hat{x}$, $U_{s}\left(
\hat{x}\right) =U_{0}\sin ^{2}k\hat{x}$,$U_{cs}(\hat{x})=U_{0}\sin k\hat{%
x}\cos k\hat{x}$ and $U_{0}\simeq g^{2}/\Delta _{a}$. We arrive at
an
effective master equation for the system:%
\begin{equation}
\dot{\rho}=-\frac{i}{\hbar }\left[ H,\rho \right] +L_{c}\rho +L\rho ,
\label{master}
\end{equation}%
where the Hamiltonian evolution is governed by

\begin{equation}
H=\frac{\hat{p}^{2}}{2m}-\hbar \Delta \left( a_{c}^{\dagger
}a_{c}+a_{s}^{\dagger }a_{s}\right) -\hbar U(\hat{x})+i\hbar \left( \eta
a_{c}^{\dagger }-\eta ^{\ast }a_{c}\right) ,  \label{ham}
\end{equation}%
\newline
and the Liouvilian describes dissipation via cavity decay $L_{c}\rho
=\kappa D\left[ a_{c}\right] \rho $ and $L\rho =\kappa D\left[
a\right] \rho $, where
the superoperator's $D$ action on a generic operator $c$ is defined as $D%
\left[ c\right] =2c\rho c^{\dagger }-c^{\dagger }c\rho -\rho
c^{\dagger }c$. Notice that we have ignored spontaneous emission in
the Liouvillian which amounts to two effects: decay of the excited
level to the vacuum modes and subsequent momentum diffusion of the
particle owing to random kicks produced by spontaneously emitted
photons. The validity of our treatment will be analyzed in more
detail in Secs.~\ref{simulation} and ~\ref{discussion} while here we
make the observation that the scaling of the rates of these
detrimental processes with $\Delta _{a}^{-2}$ makes them completely
negligible for the typically huge detunings that molecules are
driven at.

We will in the following use the quantum Langevin equations approach
\begin{subequations}
\label{lang}
\begin{eqnarray}
\frac{da_{c}(t)}{dt} &=&\left[ -\kappa +i\Delta _{c}\left( \hat{x}\right) %
\right] a_{c}+ia_{s}U_{cs}\left( \hat{x}\right) +\eta +a_{in}^{c}, \\
\frac{da_{s}(t)}{dt} &=&\left[ -\kappa +i\Delta _{s}\left( \hat{x}\right) %
\right] a_{s}+ia_{c}U_{cs}\left( \hat{x}\right) +a_{in}^{s}, \\
\frac{d\hat{x}(t)}{dt} &=&\frac{\hat{p}}{m}, \\
\frac{d\hat{p}(t)}{dt} &=&-\hbar \frac{dU(\hat{x})}{d\hat{x}}.
\end{eqnarray}%
where the position dependent detunings are $\Delta _{c,s}\left( \hat{x}%
\right) =\Delta +U_{c,s}\left( \hat{x}\right) $ and noise operators have
vanishing correlation functions except for $\left\langle a_{in}^{c}\left(
t\right) a_{in}^{c\dagger }\left( t^{\prime }\right) \right\rangle
=\left\langle a_{in}^{s}\left( t\right) a_{in}^{s\dagger }\left( t^{\prime
}\right) \right\rangle =2\kappa \delta \left( t-t^{\prime }\right) $.

\section{Classical dynamics}

Taking the quantum average of Eqs.~\eqref{lang} we get an infinite
set of ordinary differential equations for expectation values of
operator products. In a first approximation we can simply factorize
any operator product expectation values into field and atomic
operators and ignore noise operators, so that we obtain a finite
closed set of corresponding equations for classical field modes and
particles~\cite{domokos2001semiclassical}
\end{subequations}
\begin{subequations}
\label{classical}
\begin{eqnarray}
\frac{d}{dt}\alpha _{c}(t) &=&\left[ -\kappa +i\Delta _{c}\left( x\right) %
\right] \alpha _{c}+i\alpha _{s}U_{cs}\left( x\right) +\eta , \\
\frac{d}{dt}\alpha _{s}(t) &=&\left[ -\kappa +i\Delta _{s}\left( x\right) %
\right] \alpha _{s}+i\alpha _{c}U_{cs}\left( x\right) , \\
\frac{d}{dt}x(t) &=&\frac{p}{m}, \\
\frac{d}{dt}p(t) &=&-\hbar \frac{dU(x)}{dx}.
\end{eqnarray}%
The averages of cosine and sine modes field amplitudes are
$\alpha_c$ and $\alpha_s$ respectively, while the position and
momentum expectation value are $x$ and $p$. First quantum
correlations can be included in such a treatment by factorizing only
higher order products as we will see later.

\begin{figure}[t]
\centerline{\includegraphics[width=0.48\textwidth]{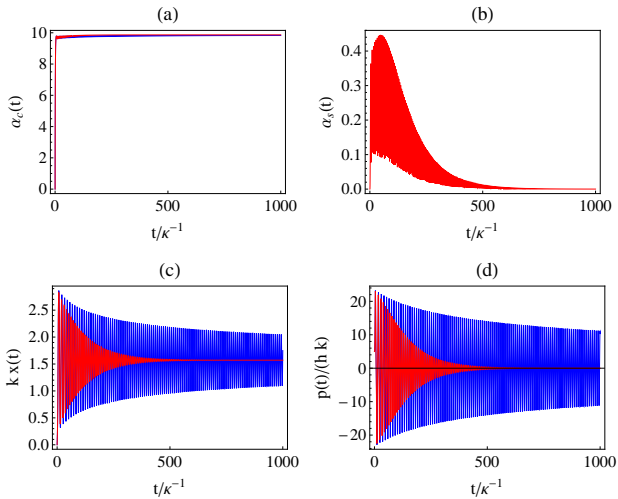}}
\caption{{}Cooling in standing wave cavity (blue) vs. ring cavity
(red). Parameters $U_{0}=0.1\times \protect\kappa $, $\Delta
_{c}=-0.3\times \protect\kappa $ and $\protect\eta =10.3\times
\protect\kappa $. (a) Field amplitude of cosine mode compared to the
standing-wave cavity mode. (b) Sine mode field amplitude. (c)
particle localization (d) momentum of particle.}
\label{ringcavcooling}
\end{figure}

Notice that the standing wave cavity model~\cite{ritsch97}can be
reproduced by the above equations by setting $U_{cs}\left( x\right)
=0$. This enables us to perform a comparison between the efficiency
of the ring cavity cooling and that of a standing wave cavity with
equal driving $\eta $. The numerical simulation results are
presented in Fig. ~\ref{ringcavcooling}. The set of parameters that
we consider $\kappa =1$ MHz, $k=2\pi \times 10^{6}$ m$^{-1}$, $m=77$
a.m.u, $U_{0}=0.1\times \kappa $, $\eta =10.3\times \kappa $ and
initial momentum $5\hbar k$ is unrealistic in that a high value of
$U_{0}$ around $100$ kHz is feasible with atoms but not with
molecules. The purpose of the exaggeration in the choice of $U_{0}$
is to illustrate the advantage of ring-cavity cooling with large
fields numerically since for smaller $U_{0}$ integration of
Eqs.~\eqref{classical} is a computationally hard and long task. The
situation presented in Fig.~\ref{ringcavcooling} is that of a
particle trapped inside a single potential well and cooled on a
timescale of order $500\times\kappa ^{-1}$ (as seen in Fig.~
\ref{ringcavcooling}d) which is much faster than the corresponding
standing-wave cavity cooling time. The detuning has been chosen at
the value of $\Delta =-0.3\times\kappa $ which is close to the
trapping frequency of the particle in the cosine mode, choice that
will be elucidated in the next Section.

\section{The strong confinement limit}

We will focus now on the quantum dynamics of this system in steady
state for strong pump and deep trapping. We drop the index of
$\alpha_c$ and denote by $\alpha $ the steady state amplitude of the
cosine mode, which we assume to be large and real (which can be
guaranteed by the proper choice of phase of $\eta $). For small
coupling $U_0 \ll \kappa$ we approximate it by $\alpha =\left\vert
\eta \right\vert /\sqrt{\kappa ^{2}+\Delta^{2}}$. Note that as
discussed in the following sections this only makes sense in the
cooling regime, where we can expect a steady state state of the
coupled system with a finite particle energy as well.

In the limit $\alpha \gg 1$, one can ignore the term  $a^{\dagger
}a\sin^{2}k\hat{x}$  in Eq. \eqref{ham} as being small compared to
the $\alpha ^{2}\cos ^{2}k\hat{x}$ term, so that the cosine mode
creates a deep periodic trapping potential. Assuming good
localization of the particle ($kx\ll 1$) at the field maxima, one
can expand  $\cos ^{2}k\hat{x}\simeq 1-k^{2}\hat{x}^{2}$ and $\sin
2k\hat{x}\simeq 2k\hat{x}$. The Hamiltonian of Eq. \eqref{ham} then
becomes
\end{subequations}
\begin{equation}
H=\left[ \frac{\hat{p}^{2}}{2m}+\frac{1}{2}m\omega _{m}^{2}\hat{x}^{2}\right]
-\hbar \Delta a^{\dagger }a-\hbar U_{0}^{\prime }(a+a^{\dagger })\hat{x},
\label{linearham}
\end{equation}%
with effective harmonic trapping frequency $\omega _{m}^{2}=2\hbar
U_{0}\left( k\alpha \right) ^{2}/m$ and rescaled effective
interaction strength $U_{0}^{\prime }=U_{0}k\alpha $. \textit{This
interaction rescaling is similar to the case of optomechanical
systems with mirrors or membranes in the limit of large photon
numbers}~\cite{Wilson-Rae-theory-07,
genes-ground-08,Marquardt-quantum-07} where a fundamentally small
coupling is collectively enhanced with the square root of the photon
number. Note that a similar collective effect could be in principle
used also for single standing wave mode cooling. However, as the
atoms then are confined close to a maximum of this mode, where its
derivative is zero we only get a coupling being second order in $x$,
which is intrinsically much smaller leading to a weaker cooling
effect.

\subsection{Fourier space analysis}
\label{fourier}
Reducing the operators to dimensionless quadratures, $\hat{Q}=\hat{x}%
/x_{zpm}$ and $\hat{P}=\hat{p}/p_{zpm}$ (where the ground state position uncertainty $%
x_{zpm}=\sqrt{\hbar /m\omega _{m}}$ and its corresponding momentum spread $%
p_{zpm}=\hbar /x_{zpm}$), one can write linearized Langevin equations in the form:
\begin{subequations}
\label{linearlang}
\begin{eqnarray}
\frac{d}{dt}a(t) &=&(i\Delta -\kappa )a(t)+i\bar{U}_{0}Q+a_{in}, \\
\frac{d}{dt}a^{\dagger }(t) &=&(-i\Delta -\kappa )a^{\dagger }(t)-i\bar{U}%
_{0}Q+a_{in}^{\dagger } \\
\frac{d}{dt}Q(t) &=&\omega _{m}P, \\
\frac{d}{dt}P(t) &=&-\omega _{m}Q+\bar{U}_{0}a+\bar{U}_{0}a^{\dagger }.
\end{eqnarray}%
The new coupling strength is
\end{subequations}
\begin{equation}
\bar{U}_{0}=x_{zpm}U_{0}^{\prime }=\left( kx_{zpm}\right) \times \left(
U_{0}\alpha \right) .  \label{uobar}
\end{equation}%
Note that the implicit dependence of $\bar{U}_{0}$ on $\alpha $
through the Lamb-Dicke parameter $kx_{zpm}\sim \omega
_{m}^{-1/2}\sim \alpha ^{-1/2}$ \textit{reduces the collective
enhancement scaling to} $\bar{U}_{0}\sim \alpha ^{1/2}$. The
difference to the typical scaling with $\alpha $ as in
optomechanical setups lies in the fact that the same field that
mediates the interaction, namely the cosine mode, is also
responsible with creating the optical trap.

One can proceed to solve Eqs.~\eqref{linearlang} as
in~\cite{Wilson-Rae-theory-07} by deriving an effective reduced
master equation for the motional degree of freedom or as
in~\cite{genes-ground-08} by analytically deriving spectra of
correlations of operators and integrating to find variances in
steady state. For example, starting from Eqs. \eqref
{linearlang} one can derive the spectrum of the $Q$ variance as $%
\left\langle Q\left( \omega \right) Q\left( \omega ^{\prime }\right)
\right\rangle =S_{Q}\left( \omega \right) \delta \left( \omega +\omega
^{\prime }\right) $ and integrate to find $\left( \Delta Q\right) ^{2}\simeq
n_{at}+1/2$. We state in the following results obtained via any of these
methods in the perturbative limit where $\bar{U}_{0}\ll \kappa $. The
physics of cooling in this regime is elucidated by considering the
scattering rates into optical sidebands%
\begin{equation*}
A\left( \omega \right) =\frac{\kappa \bar{U}_{0}^{2}}{\kappa ^{2}+\left(
\omega -\Delta \right) ^{2}}.
\end{equation*}%
An effective cooling rate is obtained as a difference between Antistokes and
Stokes rates
\begin{equation}
\Gamma =A\left( \omega _{m}\right) -A\left( -\omega _{m}\right)
\label{Gamma}
\end{equation}
and the final occupancy of
\begin{equation}
n_{at}=\frac{\kappa ^{2}+\left( \omega _{m}+\Delta \right)
^{2}}{-4\omega _{m}\Delta }
\label{natocc}
\end{equation}
is optimized to $\left( \kappa /2\omega _{m}\right) ^{2}$ under the
condition of optimal cooling $\Delta =-\omega _{m}$.

The spectrum of the field fluctuations can be easily shown to be
proportional to the spectrum of motion $ n_{a}\left( \omega \right)
= A\left( \omega \right) S_{Q}\left(\omega \right) $. Integration
leads to a simple expression:
\begin{equation}
n_{a}=\frac{A\left( \omega _{m}\right) A\left( -\omega _{m}\right) }{4\kappa \Gamma }=\frac{U_{0}}{-8\Delta }.
\label{na}
\end{equation}
The apparent divergence in the expressions of $n_{at}$ and $n_{a}$
is related to the fact that for very small detunings one leaves the
cooling regime~\cite{gangl-cold-00} and the particle spreads out
over a spatial range beyond the validity of the linearization. This
behavior shows up as well in the numerical simulations of the full
model (see Sec.~\ref{simulation}) and is more thoroughly analyzed in
Sec.~\ref{validity}, where we show that the validity of our
linearized treatment imposes that $\left\vert \Delta \right\vert $
cannot be smaller than the recoil frequency $\omega _{rec}=\hbar
k^{2}/2m$.

\subsection{Validity of tight confinement treatment}
\label{validity}
 The validity of the linearized treatment requires
localization of the particle within an optical wavelength which in a
first approximation imposes
that $kx_{zpm}\ll 1$. In other words this requires that the trap frequency $%
\omega _{m}$ be much larger than the recoil frequency $\omega
_{rec}=\hbar k^{2}/2m$,
\begin{equation*}
\omega _{rec}\ll \omega _{m}\text{.}
\end{equation*}

Moreover, localization of the particle in the final state described
by the occupancy $n_{at}$ of Eq.~\eqref{natocc} requires that the
spread of the final state
(a thermal state) is also smaller than a wavelength, i.e, $%
kx_{zpm}\sqrt{n_{at}}<1$. Assuming $\Delta $ small compared to
$\omega _{m}$ and $\omega _{m}$ of the order of $\kappa $, this
condition is equivalent to
\begin{equation*}
\left\vert \Delta \right\vert >\omega _{rec}\text{.}
\end{equation*}

As discussed in a number of papers~\cite{Wilson-Rae-theory-07,
genes-ground-08,Marquardt-quantum-07}, sideband cooling works best
when the Antistokes sideband is maximally enhanced, i.e. when
$\Delta =-\omega _{m}$.
This implicitly sets a value for $\omega _{m}$%
\begin{equation*}
\omega _{m}^{2}=\frac{1}{2}\kappa ^{2}\left[ \sqrt{1+\frac{16\omega
_{rec}U_{0}\eta ^{2}}{\kappa ^{4}}}-1\right] .
\end{equation*}%
Moreover, to resolve the sidebands one has to also ask that $\omega
_{m}>\kappa $ which is equivalent to%
\begin{equation*}
2\frac{\omega _{rec}U_{0}\eta ^{2}}{\kappa ^{4}}>1.
\end{equation*}%
One can see that this requirement is not inconsistent with the
assumption of our perturbative treatment, namely\ $\bar{U}_{0}\ll
\kappa $, which, under the assumption that $\omega _{m}$ is of the
order of $\kappa $ can be
reexpressed as%
\begin{equation*}
2\frac{\omega _{rec}U_{0}\eta ^{2}}{\kappa ^{4}}\ll \frac{\kappa }{U_{0}}\text{%
,}
\end{equation*}%
and which can be easily fulfilled for single molecules given the
smallness of the coupling $U_{0}$.

\begin{figure}[b]
\centerline{\includegraphics[width=0.50\textwidth]{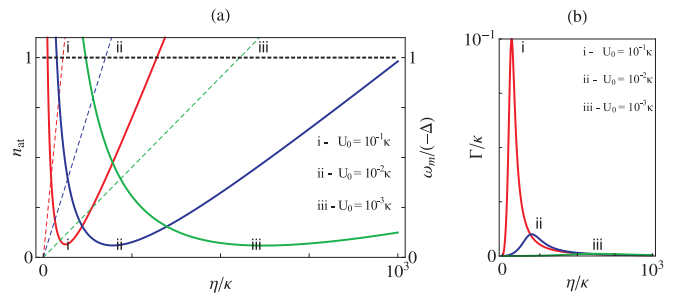}}
\caption {Atomic occupancy (a) and cooling rate (b) as a function of
increasing $\eta$ for fixed $\Delta=-2\kappa$. The minima of the
full lines (i, ii and iii) occur at the point where the 3 dashed
lines (that show $-\omega_{m}/\Delta$) intersect the dashed black
line of value 1, i.e. when the optimal sideband cooling condition is
fulfilled. The cooling rates are also optimal, as expected, at
values of $\eta$ which ensure $\omega_{m}=-\Delta$.}
\label{fixeddelta}
\end{figure}

\subsection{Coupled equations for operator expectation values}

Starting from Eq. \eqref{master} where the linear dynamics is given
by the simplified Hamiltonian of Eq. \eqref{linearham} and following
the ideas shown in~\cite{gangl1999collective} one can derive a
closed set of linear differential equations containing expectation
values of pairwise operator products only. These can be solved
dynamically and also exact steady state values for

$(\bn{a^{\dagger}a},\bn{Q^2},\bn{P^2})$ can be deduced:
\begin{subequations}
\begin{eqnarray}
\frac{d}{dt}\bn{a^\dagger a} &=& -2\kappa\bn{a^\dagger a}-i\bar
U_0(\bn{aQ}-\bn{a^\dagger Q}), \\
\frac{d}{dt}\bn{Q^2} &=& \omega_m\bn{\mathcal{A}}, \\
\frac{d}{dt}\bn{\mathcal{A}} &=& 2\omega_m(\bn{P^2}-\bn{Q^2})+\nonumber\\
                 & & 2\bar{U_0}(\bn{aQ}+\bn{a^\dagger Q}), \\
\frac{d}{dt}\bn{P^2} &=& -\omega_m\bn{\mathcal{A}}+2\bar
U_0(\bn{aP}+\bn{a^\dagger P}), \\
\frac{d}{dt}\bn{aQ} &=& \omega_m\bn{aP}-(\kappa-i\Dc)\bn{aQ}+i\bar
U_0\bn{Q^2}, \\
\frac{d}{dt}\bn{aP} &=& (-\kappa+i\Dc)\bn{aP}-\omega_m\bn{aQ}+\nonumber\\
             & &  \bar
U_0\left(\bn{a^{\dagger}a}+1/2+\bn{a^2}+\frac{i}{2}\bn{\mathcal A}\right), \\
\frac{d}{dt}\bn{a^2} &=& -2(\kappa-i\Dc)\bn{a^2}+2i\bar U_0\bn{aQ},
\end{eqnarray}
\end{subequations}%
where $\bn{\mathcal{A}}=\bn{QP+PQ}$. One can therefore derive
steady-state solutions from this set of coupled equations for the
motional quadratures and intracavity photon number
\begin{equation*}
 \bn{a^{\dagger}a} =
-\frac{\bar
U_0^2(\Dc^2+\kappa^2)}{4\omega_m\Dc(\kappa^2+\Dc^2)+8\bar
U_0^2\Dc^2},
\end{equation*}

\begin{equation}
\bn{Q^2} = -\frac{(\kappa^2+\omega_m^2+\Dc^2)(\kappa^2+\Dc^2)+2\bar
U_0^2\omega_m\Dc}{4\omega_m\Dc(\kappa^2+\Dc^2)+8\bar U_0^2\Dc^2}
\end{equation}

\begin{equation}
\bn{P^2}= -\frac{(\kappa^2+\omega_m^2+\Dc^2+2\bar
U_0^2\Dc/\omega_m)(\kappa^2+\Dc^2)+2\bar
U_0^2\omega_m\Dc}{4\omega_m\Dc(\kappa^2+\Dc^2)+8\bar U_0^2\Dc^2}
\end{equation}
which in the limit where $\bar U_0\ll\kappa$ are approximated by the
expressions listed in Sec.~\ref{fourier}. The method is a little
more laborious, but instructive since it provides one with results
that hold for arbitrary coupling strength $U_0$ and with some
insight into the physics of the cooling process as seen from the
relation derived for the intracavity photon number
\begin{equation*}
\bn{a^{\dagger}a} = -\frac{\omega_m}{2\Dc}(\bn{Q^2}-\bn{P^2}).
\end{equation*}
This shows that $n_{a}$ is directly proportional to the imbalance
between the uncertainties $\bn{Q^2}$ and $\bn{P^2}$, thus giving a
good measure of the 'squashing ' of the uncertainty disk
characteristic of a coherent state (or vacuum).

\begin{figure}[t]
\centerline{\includegraphics[width=0.50\textwidth]{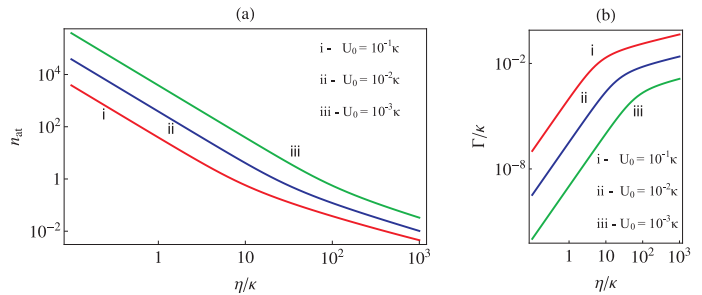}}
\caption {Log-log plots of atomic occupancy (a) and cooling rate (b)
as a function of increasing $\eta$ for variable $\Delta$ that is
dynamically changed to fulfill the optimal cooling condition
$\omega_{m}=-\Delta$. The occupancy goes to 0 as expected for large
$\eta$ as the system goes deep into the resolved sideband regime (as
seen in (a)). The cooling rate increases at first with $\eta^3$ and
after entering the resolved sideband regime, its scaling drops to
$\sqrt\eta$. The illustrations are for 3 values of $U_0$ and clearly
show that small final occupancies can still be obtained even with
small $U_0$, however, at the expense of drastic increases in the
cooling times.} \label{vardelta}
\end{figure}

Moreover, the exact expressions of $\bn{Q^2}$ and $\bn{P^2}$ allow
us to make a clear investigation of the feasibility of our molecule
cooling scheme, i.e., into the cooling occupancy and rates with
diminishing $\bar U_0$. As seen in Fig.~\ref{fixeddelta}, for a
fixed $\Delta=-2\kappa$, one can increase the cavity input power to
reach a point of optimal cooling where the occupancy is minimized as
the cooling rate is maximized, and which is given by the condition
$\omega_{m}=-\Delta$. This is valid for any arbitrarily small
$U_{0}$. However, as $U_{0}$ is set to smaller values, the cooling
rates are becoming unrealistically small. Setting the optimal
cooling condition, $\omega_{m}=-\Delta$, the results of Fig.
~\ref{vardelta}(a) are straightforward, showing that the occupancy
asymptotically tends to 0 with increasing $\eta$. Investigating the
analytical expression for $\Gamma$ as given in Eq. ~\eqref{Gamma},
we see that in the unresolved sideband regime where
$\omega_m\ll\kappa$, the scaling is with $\eta^3$ while deep in the
resolved sideband regime where $\kappa\ll\omega_m$ the scaling drops
to $\sqrt\eta$. This is clearly illustrated in
Fig.~\ref{vardelta}(b) where the turning point occurs around an
$\eta$ for which $\omega_m \approx \kappa$.

\section{Numerical simulations}\label{simulation}

\begin{figure*}[t]
\centerline{\includegraphics[width=0.75\textwidth]{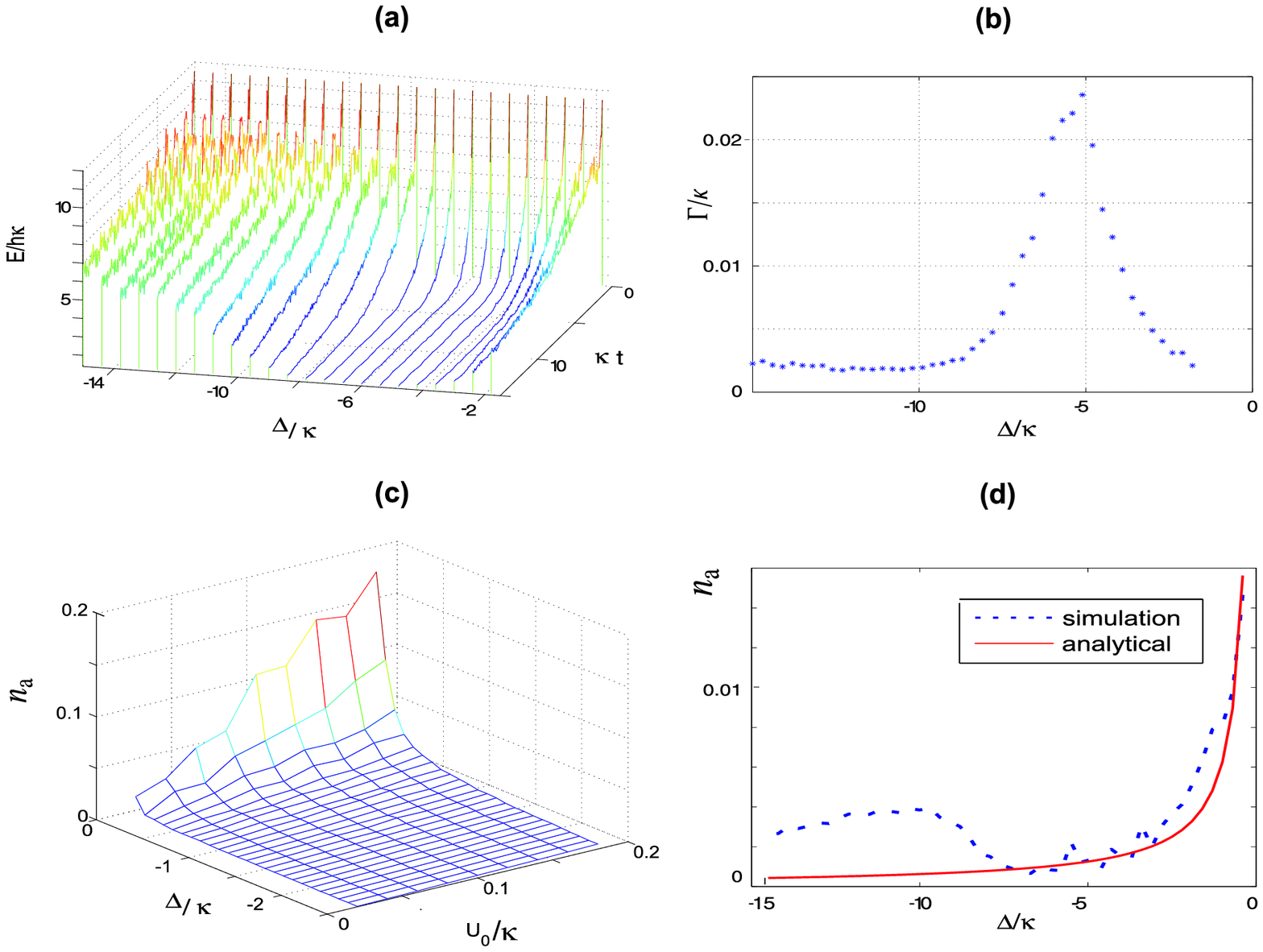}}
\caption{For a), b) and c), parameters are $U_{0}=0.01\times
\protect\kappa $ and $\omega_m= 6\kappa$. a) Time dependence of
average kinetic energy (mean over 100 trajectories) of the particle
in units of $\hbar \kappa$ as a function of detuning. b) Cooling
rate as a function of detuning. c) Simulated time averaged photon
number in steady state as a function of detuning and coupling
strength. d) Simulated and  theoretical average sine mode photon
number in steady state calculated in the linearized model as a
function of detuning for $U_0=\kappa/20$.} \label{Fig5}
\end{figure*}

 After having demonstrated the key physical
mechanisms present in the dynamics using different approximative
analytical treatments, we will now check the range of validity of
these results by a numerical solution of the full quantum model. The
quantum dynamics can be efficiently approximated using time
dependent Monte Carlo wavefunction simulations of the full coupled
atom field system. Here we can conveniently make use of our
established previously developed C++QED simulation
package~\cite{C++qed,vukics2007c++}. Due to symmetry we can limit
position space to one optical wavelength and truncate the photon
numbers of the field modes sufficiently above the expected mean
values, so that the overall size of the Hilbert space stays easily
tractable. For very large pump photon numbers, the cosine mode
dynamics can be approximated by a coherent field, so that only two
quantum degrees of freedom have to be explicitly treated. This
allows for even faster simulation and averaging over a large number
of trajectories.

\subsection{Cooling rates and steady state temperature}

\begin{figure*}[t]
\centerline{\includegraphics[width=0.75\textwidth]{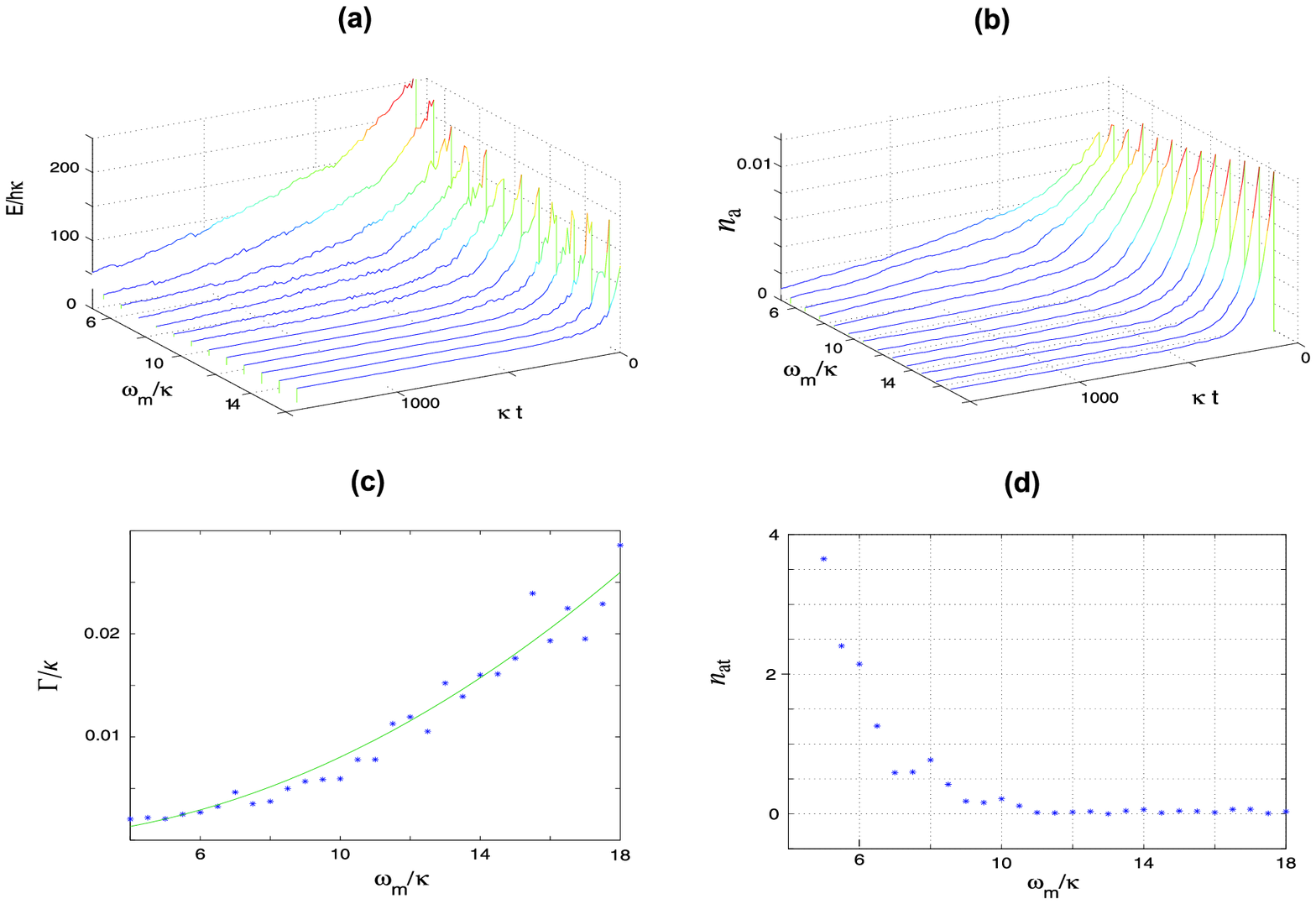}}
\caption{Fixed parameters are $U_0=\kappa/400$ and
$\Delta=-\omega_m$. a) Averaged particle kinetic energy over 500
trajectories as a function of time for increasing trapping frequency
starting with initial kinetic energy of $E_k\approx 25\hbar \kappa$.
b)Average steady state photon number in sine-mode. We see the
decrease associated with the better localization in a deeper
trapping potential. c) Dots give the cooling rate $\gamma$ averaged
over 500 trajectories. The solid line shows the expected power
dependence derived from the linearized model above. d) Final steady
state oscillator occupation number after a time $\kappa t = 300$. }
\label{Fig6}
\end{figure*}

As a first step we confirm the efficiency of cooling as a function
of cavity detuning by starting with a particle in a highly excited
state and calculating its average energy loss as function of time
for varying detuning between pump and cavity resonance. The kinetic
energy as function of time is shown in Fig.~\ref{Fig5}(a). Close to
optimal detuning  $\Delta$ of we see a fast decay towards the ground
state kinetic energy $\hbar \omega_m/4$ with an optimal value of the
detuning near $\Delta=-\omega_m$. This value is different from the
optimal value for $\Delta \approx -\kappa$ obtained for weak field
cavity cooling and is characteristic for the optomechanical cooling
limit where the mechanical oscillation frequency of the particle is
bigger than the cavity linewidth $\kappa$. Less cooling or even
heating occurs for small detunings in contrast with the
optomechanical model predictions; the reason is the breakdown of the
tight confinement model as explained in Sec.~ \ref{validity}. A
central important quantity for the usefulness of this cooling method
is the corresponding cooling rate, which we plot in
Fig.~\ref{Fig5}(b). It also peaks near the sideband frequency
$\omega_m$ and reaches almost the order of magnitude of $\kappa$
well above $U_0$, which gives a practically quite useful timescale.

Let us mention here that the simulations also give the steady state energy as
a function of coupling strength $U_0$ and detuning. We find that for
sufficiently large detuning we get ground state cooling also for
very small coupling $U_0$. However, as expected for smaller $U_0$
the timescale to reach the ground state gets longer. This is
different from a standard optomechanics setup, where the mirror is
additionally coupled to a thermal reservoir and the cooling rate
itself largely determines the final temperature. The corresponding
simulation results are in good agreement with the analytic results
obtained above.

Here we will show another aspect of this. As the cooling is
connected to scattering of photons into the sine mode, its photon
number $n_{a}$ directly provides information on the particle's state
and in particular on its temperature. The dependence of the
sine-mode photon number $n_{a}$ on detuning and coupling strength is
shown in Fig.~\ref{Fig5}(c). It shows the predicted divergence at
low detunings and agrees fairly well with the theoretical prediction
from the linearized model above (See Fig.~\ref{Fig5}(d)), as in
steady state the particle is well localized and linearization is
justified.

Let us note that that the sine-mode scattered photon number, i.e.
$n_{a}$ is independent of mass, which is surprising according to a
simple logic that a better localized particle would be expected to
scatter less into the sine-mode, which is zero at its trapping
point. We now add a direct comparison of the simulated and
theoretically predicted average photon number in Fig.~\ref{Fig5}(d).
While for smaller detunings the agreement is excellent, we see that
at larger detunings the simulation predicts higher numbers. To a
large extend this reflects the fact that even after a simulation
time of $\omega_{rec}t=1000$ the system has not reached its steady
state value.

As next important issue we confirm the possibility of fast cooling
even for very small coupling. For this purpose we show how the
initial kinetic energy of a particle decays for small effective
coupling $U_0=\kappa/400$ with increasing pump power. A deeper
potential leads to higher trapping frequencies, faster cooling and
better final particle localization. This is visible in
Fig.~\ref{Fig6}(a), where we plot the loss of kinetic energy of a
trapped particle as an average over 500 trajectories. We see that
for the same initial kinetic energy we get much faster relaxation
with higher trapping frequency. This behavior gets even more
obvious, if one looks at the average occupation number of the
trapped particle as function of time. While the final kinetic energy
first drops due to better cooling but then increases with higher
$\omega_m$ due to the increasing zero point energy of the particle,
the average particle vibrational excitation in the trap continuously
drops and reaches almost zero for deep traps. Of course in practise,
at some point one gets limited by the very small but yet nonzero
spontaneous emission rate, which will induce heating and particle
loss on a very long time scale.

The cooling can be directly and nondestructively monitored in the
scattered photon number shown Fig.~\ref{Fig6}(b) which directly
reflects better ground state cooling with increasing trap depth. The
detailed correspondence of the two quantities as calculated in the
analytical section can be even more clearly seen in individual
trajectories as discussed in the next section.

One central question is the efficiency of the cooling for very far
off resonance excitation, where the coupling of the particle to the
field modes gets small. Here we can make use of very high pump mode
photon numbers as in optomechanics. Extracting the corresponding
decay rate from an exponential fit in Fig.~\ref{Fig6}(c) we see its
nonlinear increase with $\omega_m$ in accordance with the expected
behavior form the linearized model.

To show that this faster cooling does not enhance the diffusion
significantly, we also plot the long time average of the oscillator
excitation number in Fig.~\ref{Fig6}(d), which predicts ground state
cooling at large $\omega_m$.

As a bottom line we clearly see that with sufficient power in the
cosine mode to generate large mechanical trapping frequencies we can
get fast molecular cooling despite very small coupling. As the
induced photon number in the sine mode still stays small, we expect
that this result holds even for more particles cooled simultaneously
as long the photon number in the sine mode does not increase too
much. We plan to study this scaling in more detail in the future.

\begin{figure}[t]
\centerline{\includegraphics[width=0.46\textwidth]{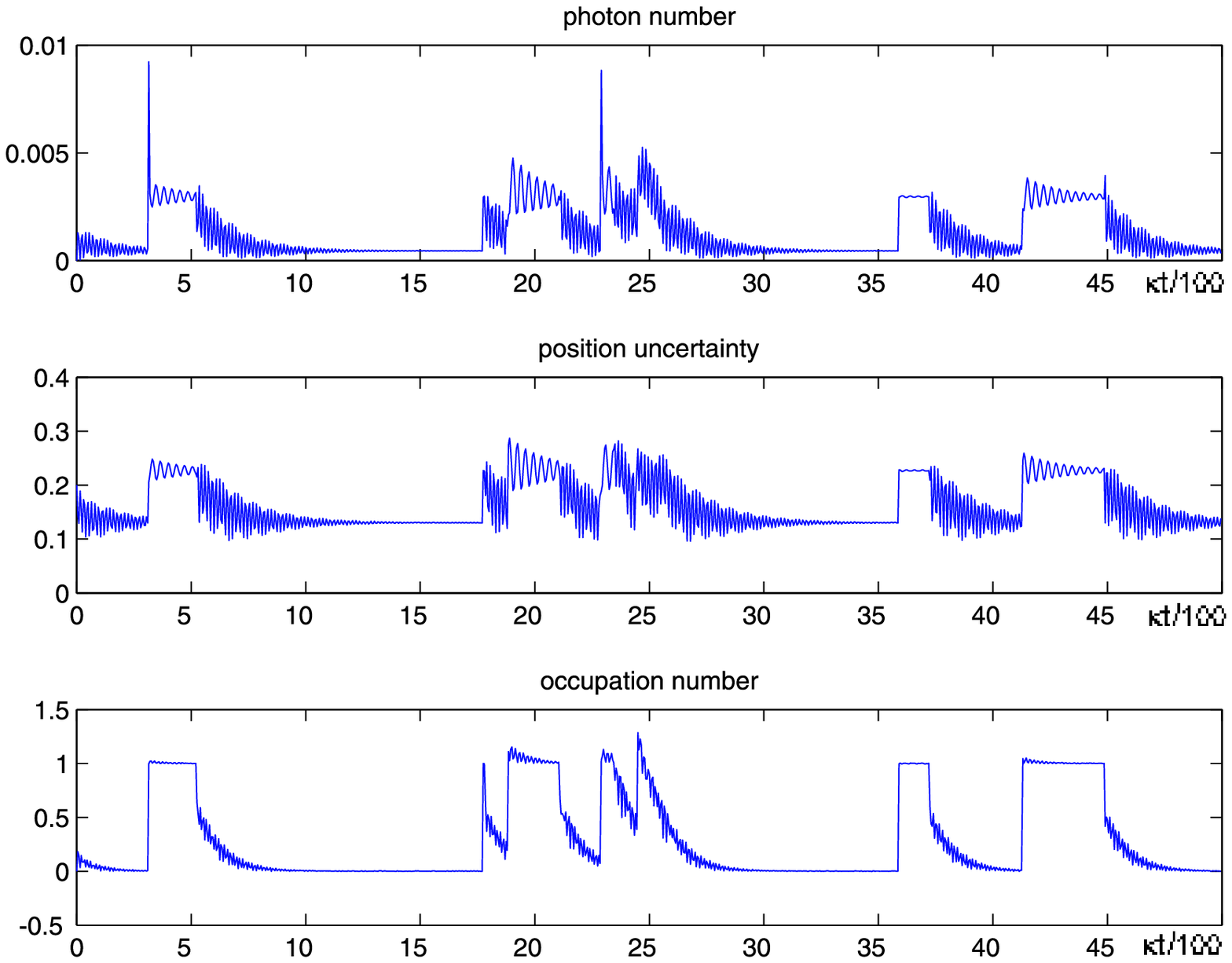}}
\caption{Single trajectory expectation values showing the particle
occupation number, the photon number in the sine mode and the
particles position uncertainty as a function of time for  $\Delta =
- 2\kappa$ and other parameters as in Fig.~\ref{Fig5}.}
\label{quantumjumps}
\end{figure}

\subsection{Quantum Jumps}
One very useful advantage of the Monte Carlo simulation approach is
that besides averages and expectation values one can also look at
individual trajectories, from which one can get interesting clues on
the underlying dynamics and direct hints for the expected
experimental data. In Fig.~\ref{quantumjumps} we show a time window
of the results of a typical trajectory as used above to find
averages. In order to exhibit interesting dynamics we have chosen
the case of a bit lower than optimum detuning $\Delta = -2\kappa <
\omega_m$ still yielding a non-vanishing occupation of the first
excited trap level. In the simulation we continuously monitor the
photons leaving the sine-mode. As we always know the momentary
atom-field wavefunction, we can easily show its correlation with any
other system variable.

As expected we see a clear correspondence between the time evolution
of the atomic position uncertainty and occupation number and the
average photon number in the sine mode. Note that after reaching the
ground state the system shows striking and clear quantum jumps
between the lowest and first excited state. While it spends long
periods in the ground state we see sporadic jumps to the first
excited state where the particle remains for surprisingly long time
until jumping back. This should be related with a symmetry change in
the particle wave function during such a jump. While such jumps are
forbidden in a standing wave cavity setup, they attain a finite
probability in a ring cavity~\cite{horak-dissipative-01}, where
spontaneous shifts of the trap minimum allow to break the mirror
symmetry of the trap. As these fluctuations are rather rare, the
first excited state, once excited, is much more long lived than the
second excited trap state, which possesses the same symmetry as the
ground state. The observation of such quantum jumps thus would
clearly reveal the quantum nature of the molecule's center of mass
dynamics.

\section{Discussion}
\label{discussion} Lets now add some remarks on the effects of the
spontaneous photon scattering rate $\gamma $. Naturally a molecule
is not a simple two-level system. Nevertheless considering a single
ground state and one manifold of excited states centered around
$\omega _{a}$ with an energy spread bandwidth smaller than the laser
detuning $\Delta _{a}\,$, can well describe the induced polarization
dynamics. For weak saturation a two-level approach with an effective
dipole matrix element is then sufficient. In the limit of
$\Delta_{a}\gg \gamma $, the induced molecular polarization can be
decomposed into real and imaginary parts describing dispersion
(proportional to $U_{0}$) and absorption (proportional to $\gamma
_{0}=\gamma g^{2}/\Delta _{a}^{2}$). To lowest order, $\gamma _{0}$
supplements the coupling strength $\bar{U}_{0}$ in
Eqs.~\eqref{linearlang} with an imaginary part
$i\bar{\gamma}_{0}=i\left( kx_{zpm}\right) \times \left( \gamma
_{0}\alpha \right) $. Within our model, this implies a renormalizing
of the
coupling strength from $\bar{U}_{0}$ to $\sqrt{\bar{U}_{0}^{2}+\bar{\gamma}%
_{0}^{2}}$. In view of $\bar{\gamma}_{0}\ll \bar{U}_{0}$, the effect is
negligible.

A second effect is the diffusion of the momentum triggered by
the spontaneously emitted photons. A rate proportional to $\bar{\gamma}%
_{0}\alpha ^{2}\left( kx_{zpm}\right) ^{2}$ up to a geometrical factor is
obtained that can again be neglected based on the smallness of $\bar{\gamma}%
_{0}$ and the inhibition with the square of the Lamb-Dicke parameter.

One can extend Eqs.~\eqref{linearlang} to also include quantum
fluctuations of the cosine mode. The fluctuations of this mode are
close to those of the vacuum state with extra terms coming from the
scattering of the sine mode photons off the molecule. Qualitatively
this leads to a fluctuating trapping frequency $\omega _{m}$ (that
is
proportional to $\alpha ^{2}$) and a fluctuating coupling strength $\bar{U}%
_{0}$. For a quantitative analysis, one can notice that in the
regime which is optimal for cooling where $\Delta =-\omega _{m}$,
the sine mode population is always close to zero (as shown in
Eq.~\eqref{na}) and backscattering into the cosine mode can be
neglected. When operating close to resonance, surely a large $U_{0}$
can lead to a considerable number of photons in the sine mode and
the back action and potential shift effects listed above have to be
taken into account, for example by following an approach similar to
the one in~\cite{rabl-phase-09}.

As first noticed in~\cite{vitali-optomechanical-07}, the linear
optomechanical interaction described by Eqs.~\eqref{linearlang} can
lead to entanglement between the two degrees of freedom: vibration
of molecule and quadratures of the sine mode, owing to
quadrature-quadrature interaction. In terms of the logarithmic
negativity, a maximum value of around $0.4$ around the optimal
cooling regime can be obtained and can be even surpassed by a proper
filtering of the cavity output field as suggested
in~\cite{genes-robust-08}. As an advantage over the mirror-field
entanglement scheme which is strongly affected by the thermal bath,
the molecule-field entanglement is effectively obtained under zero
temperature conditions.

\section{Conclusions}

In conclusion we have shown that the ring cavity setup allows for
strong coupling between a molecule's motion and the sine mode via
scattering of sideband photons from the cosine mode into the sine
mode. The low single molecule field coupling is enhanced by the
large photon number of the strongly driven cosine mode. In this
regime the dynamics is similar to the one described in typical
optomechanical systems with mirrors or membranes and we predict fast
cooling rates. In contrast to superradiant cooling
\cite{chan2003observation,salzburger2009collective} we do not need
large molecular ensembles from start but rely on a single particle
cooling mechanism. The central technical challenge here seems to
generate high enough intensities in a ring cavity to assure trap
frequencies larger than the cavity linewidth even at very large pump
detunings from any optical resonance. Here mirrors with a finesse
over $F=1000$ and supporting megawatt intracavity power are needed.
From the molecular side one would like to have a high polarizability
per mass with very low absorption at the chosen operating frequency.
In this case even ground state cooling of any polarizable object
seems feasible in a suitable ring cavity. As for atoms it could also
be hoped that cooling properties get even more favorable in cavities
invoking even more modes as in a confocal setup~\cite{cavcoolrev}.

\textit{Acknowledgements - } We acknowledge support from the EC (FP6
Integrated Project QAP, and FET-Open project MINOS) and Euroquam Austrian
Science Fund project I119 N16 CMMC.


\begin{thebibliography}{99}
\bibitem{C++qed} This object oriented C++ package is freely available and documented at
sourceforge.net.
\end{thebibliography}
\end{document}